\documentclass[amsmath,amssymb,superscriptaddress,nobalancelastpage,prl,twocolumn,preprintnumbers]{revtex4-1}

\usepackage{bm}
\usepackage{braket}
\usepackage{diagbox}
\usepackage{graphicx}
\usepackage{booktabs}
\usepackage{hyperref}
\hypersetup{colorlinks,linkcolor=blue,urlcolor=blue,citecolor=blue}
\usepackage{siunitx}
\usepackage{ulem}
\usepackage{varioref}
\usepackage{xr-hyper}
\usepackage{xcolor}

\newcommand{\PNO}{PrNiO$_2$}
\newcommand{\SCO}{SrCuO$_2$}

\begin{document}

\author{Xunyang~Hong}
\thanks{These authors contributed equally to this work.}
\affiliation{Department of Physics, The Chinese
University of Hong Kong, Shatin, Hong Kong, China}
\affiliation{Physik-Institut, Universit\"{a}t Z\"{u}rich, Winterthurerstrasse 190, CH-8057 Z\"{u}rich, Switzerland}

\author{Yuetong~Wu}
\thanks{These authors contributed equally to this work.}
\affiliation{Department of Physics, The Chinese
University of Hong Kong, Shatin, Hong Kong, China}
\affiliation{State Key Laboratory of Quantum Information Technologies and Materials, The Chinese University of Hong Kong, Shatin, Hong Kong, China}

\author{Ying~Chan}
\affiliation{Department of Physics, The Chinese
University of Hong Kong, Shatin, Hong Kong, China}
\affiliation{State Key Laboratory of Quantum Information Technologies and Materials, The Chinese University of Hong Kong, Shatin, Hong Kong, China}

\author{Sze Tung~Li}
\affiliation{Department of Physics, The Chinese University of Hong Kong, Shatin, Hong Kong, China}

\author{I.~Bia\l{}o}
\affiliation{Physik-Institut, Universit\"{a}t Z\"{u}rich, Winterthurerstrasse 
190, CH-8057 Z\"{u}rich, Switzerland}

\author{L.~Martinelli}
\affiliation{Physik-Institut, Universit\"{a}t Z\"{u}rich, Winterthurerstrasse 190, CH-8057 Z\"{u}rich, Switzerland}

\author{A.~Drewanowski}
\affiliation{Physik-Institut, Universit\"{a}t Z\"{u}rich, Winterthurerstrasse 190, CH-8057 Z\"{u}rich, Switzerland}

\author{Qiang Gao}
\affiliation{Beijing National Laboratory for Condensed Matter Physics, Institute of Physics, Chinese Academy of Sciences, Beijing 100190, China}

\author{Xiaolin Ren}
\affiliation{Beijing National Laboratory for Condensed Matter Physics, Institute of Physics, Chinese Academy of Sciences, Beijing 100190, China}

\author{Xingjiang Zhou}
\affiliation{Beijing National Laboratory for Condensed Matter Physics, Institute of Physics, Chinese Academy of Sciences, Beijing 100190, China}

\author{Zhihai Zhu}
\affiliation{Beijing National Laboratory for Condensed Matter Physics, Institute of Physics, Chinese Academy of Sciences, Beijing 100190, China}

\author{A.~Galdi}
\affiliation{Dipartimento di Ingegneria Industriale, Universita' degli Studi di Salerno, Fisciano (SA) 84084, Italy}
\affiliation{Department of Materials Science and Engineering, Cornell University, Ithaca, New York 14850, USA}

\author{D.~G.~Schlom}
\affiliation{Department of Materials Science and Engineering, Cornell University, Ithaca, New York 14850, USA}
\affiliation{Kavli Institute at Cornell for Nanoscale Science, Ithaca, New York 14853, USA}

\author{K.~M.~Shen}
\affiliation{Kavli Institute at Cornell for Nanoscale Science, Ithaca, New York 14853, USA}
\affiliation{Department of Physics, Laboratory of Atomic and Solid State Physics, Cornell University, Ithaca, New York 14853, USA}

\author{J. Choi}
\affiliation{Diamond Light Source, Harwell Campus, Didcot OX11 0DE, United Kingdom}

\author{M.~Garcia-Fernandez}
\affiliation{Diamond Light Source, Harwell Campus, Didcot OX11 0DE, United Kingdom}

\author{Ke-Jin~Zhou}
\affiliation{Diamond Light Source, Harwell Campus, Didcot OX11 0DE, United Kingdom}

\author{N.~B.~Brookes}
\affiliation{European Synchrotron Radiation Facility, 71 Avenue des Martyrs, 38043 Grenoble, France}

\author{H.~M.~R{\o}nnow}
\affiliation{Institute of Physics, \'{E}cole Polytechnique Fed\'{e}rale de Lausanne (EPFL), CH-1015 Lausanne, Switzerland}

\author{Qisi~Wang}
\email{qwang@cuhk.edu.hk}
\affiliation{Department of Physics, The Chinese
University of Hong Kong, Shatin, Hong Kong, China}
\affiliation{State Key Laboratory of Quantum Information Technologies and Materials, The Chinese University of Hong Kong, Shatin, Hong Kong, China}
   
\author{J.~Chang}
\affiliation{Physik-Institut, Universit\"{a}t Z\"{u}rich, Winterthurerstrasse 190, CH-8057 Z\"{u}rich, Switzerland}


\title{Impact of Electron Correlations on Infinite-Layer Cuprates and Nickelates}

\maketitle

\textbf{Optimization of unconventional superconductivity involves a balance of interaction strengths.
Precise determination of correlation strength across different material families is therefore important. Here, we present a combined X-ray absorption spectroscopy (XAS) and resonant inelastic X-ray scattering (RIXS) study of infinite-layer \PNO\ and \SCO\ that enables fair comparison of their interaction strengths. 
For both compounds, we study the orbital and magnetic excitations and extract their dispersions along high-symmetry directions. Using a single-band Hubbard model and including higher-order exchange interactions, we derive the correlation factor $U/t$ for both compounds. A key finding is that despite a smaller Coulomb repulsion $U$, \PNO\ exhibits a correlation strength that is 20\% stronger than that of its isostructural cuprate counterpart \SCO.
This indicates that a moderation of the correlation strength may further optimize superconductivity in nickelates.}\\[2.5mm]

The discovery of superconductivity in doped and undoped infinite-layer nickelates has stimulated new research lines~\cite{li_superconductivity_2019,parzyck_superconductivity_2025,sahib_superconductivity_2025}. One important question is whether there exists a mapping between cuprate and nickelate superconductivity. Similar orbital motives exist in both compound families. 
Like cuprates, infinite-layer nickelates exhibit a layered structure with a $3d^9$ electronic configuration and strong electronic correlations~\cite{anisimov_electronic_1999,botana_similarities_2020}. However, key distinctions emerge, particularly in their electronic structure~\cite{goodge_doping_2021} and magnetic properties, posing significant challenges to our understanding~\cite{ghiringhelli_noticeable_2024}.
One critical aspect that may distinguish the underlying physics between nickelates and cuprates is the strength of electron correlations, characterized by the ratio between the on-site Coulomb repulsion $U$ to the nearest-neighbour hopping integral $t$ (i.e. $U/t$).
Within the framework of the single-band Hubbard model, which is widely used to describe correlated behaviors in both nickelates and cuprates, $U/t$ directly governs the balance between magnetic fluctuations and superconductivity~\cite{kitatani_optmizing_2023,worm_spin_2024}.

\begin{figure*}
    \centering
    \includegraphics[width=\linewidth]{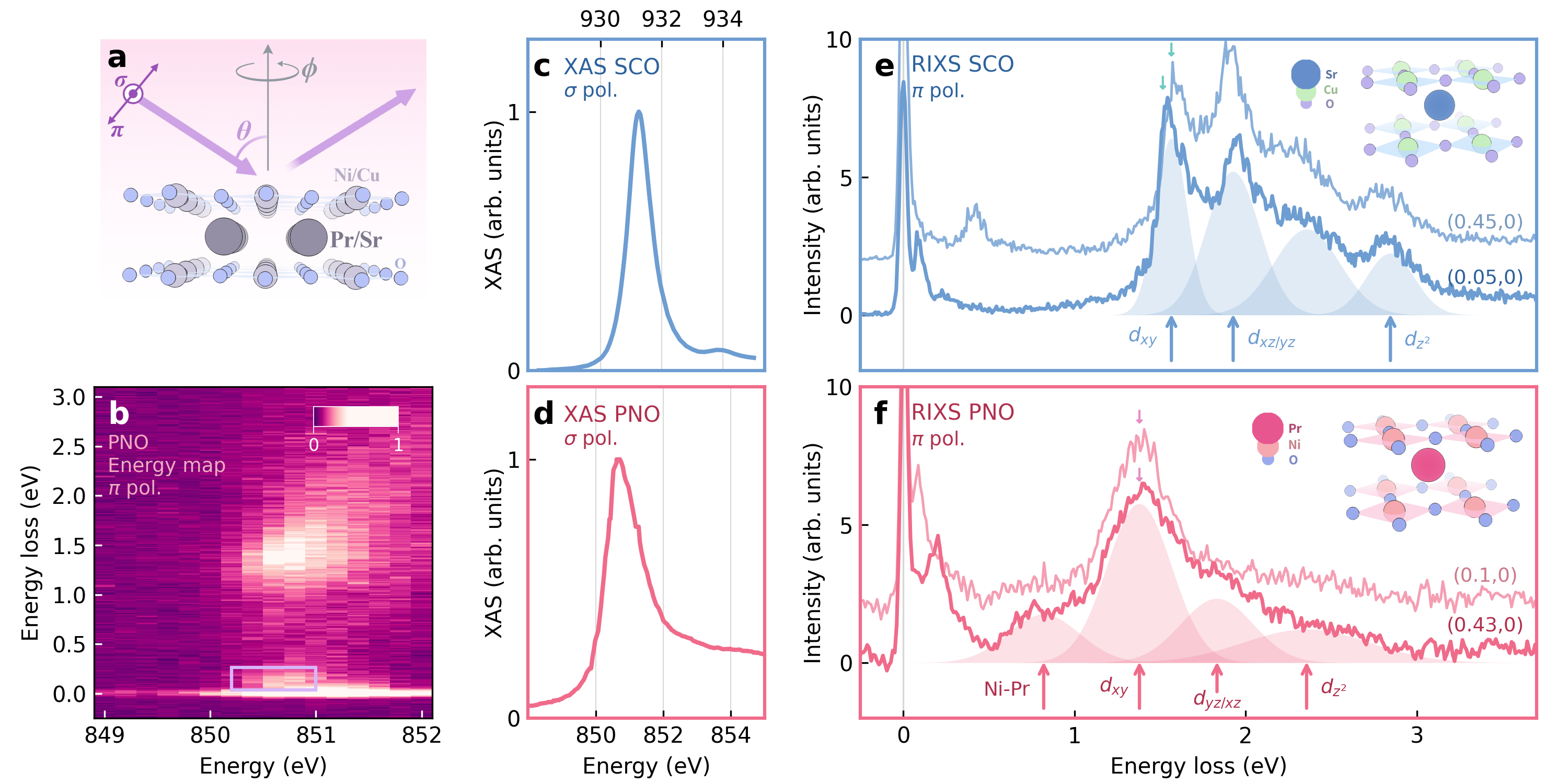}
    \caption{X-ray absorption spectroscopy (XAS) and resonant inelastic X-ray scattering (RIXS) on SrCuO$_2$ and PrNiO$_2$ films. (a) Schematic illustration of the RIXS scattering geometry --- with $\theta$ describing the incident angle and $\phi$ the sample rotation. $\sigma$ and $\pi$ indicate respectively linear horizontal and vertical incident light polarizations. (b) RIXS energy map recorded on PrNiO$_2$. The purple rectangle inset marks  the resonant magnon intensity.  (c,d) Cu $L$-resonance of SrCuO$_2$ and Ni  $L$-resonance of PrNiO$_2$ probed with XAS (e,f) RIXS spectra recorded on SrCuO$_2$ and PrNiO$_2$ at respectively the Cu and Ni $L$-edges and at in-plane momentum transfers as indicated. Small down-pointing arrows mark the shift (marginal-shift) of $d_{xy}$ excitations in \SCO\ (\PNO). Modeling of the crystal field excitations are indicated with shaded Gaussian profiles. The insets illustrate the equivalent crystal structure of SrCuO$_2$ and PrNiO$_2$.}
    \label{fig:enter-label}
\end{figure*}

Experimental attempts have been made to characterize the electronic structure and correlation effects in infinite-layer nickelates. These include angle-resolved photoemission spectroscopy (ARPES)~\cite{ding_cuprate_2024,sun_electronic_2025,li_observation_2025}, X-ray absorption spectroscopy (XAS)~\cite{xiao_superconductivity_2024}, X-ray photoemission spectroscopy (XPS)~\cite{chen_matter_2022}, resonant inelastic X-ray scattering (RIXS)~\cite{hepting_electronic_2020,lu_magnetic_2021}, and electron energy-loss spectroscopy (EELS)~\cite{goodge_doping_2021,hu_atomic_2024}.
Yet, a quantitative conclusion remains elusive due to different types of experimental limitations.
The study of collective spin excitations offers an alternative approach, as the spin exchange couplings originate from electronic correlation effects.
A recent RIXS study compared magnons in NdNiO$_2$ and YBa$_2$Cu$_3$O$_{7-\delta}$~\cite{rosa_spin_2024}. 
Using a single-band Hubbard model, this work concluded that the Coulomb repulsion $U$ of NdNiO$_2$ is twice that of YBa$_2$Cu$_3$O$_{7-\delta}$, leading to a 60\% larger $U/t$ in the nickelates.
However, this methodology has some shortcomings. First, recent ARPES results have revealed a circular Fermi surface~\cite{ding_cuprate_2024,sun_electronic_2025,li_observation_2025}, suggesting that next-nearest-neighbor hopping ($t'$) plays a significant role. Second, the compared nickelate and cuprate compounds were not isostructural. The presence of apical oxygen in cuprates is known to suppress longer-range hopping and exchange processes~\cite{peng_influence_2017,wang_magnon_2018}, which could further distort a fair comparison. 

To facilitate a  direct comparison, we  conducted a resonant inelastic X-ray scattering (RIXS) study on PrNiO$_2$, mapping out its electronic and magnetic excitations across high-symmetry directions. Our data extend beyond the scope of previous studies, providing a comprehensive characterization of the $k$-space properties of magnetic and orbital excitations. By incorporating higher-order hopping terms $t^\prime$, and  
$t^{\prime\prime}$ into our analysis, we extract refined values for $U$, and $t$ from our experimental data. These results also allow us to characterize charge fluctuations and staggered magnetization with greater accuracy.
We further compare these results with those obtained from the isostructural infinite-layer cuprate SrCuO$_2$ thin film, analyzed with the same approach. 
This direct comparison reveals a 20\% stronger correlation strength $U/t$ in \PNO\ compared to \SCO. Our analysis of the orbital excitations further indicates a reduced orbital exchange in \PNO\ resulting from stronger electronic localization. Implications for electronic properties such as staggered magnetic moments, superconductivity, and strange metal properties are discussed.



\begin{figure*}
    \centering
    \includegraphics[width=\linewidth]{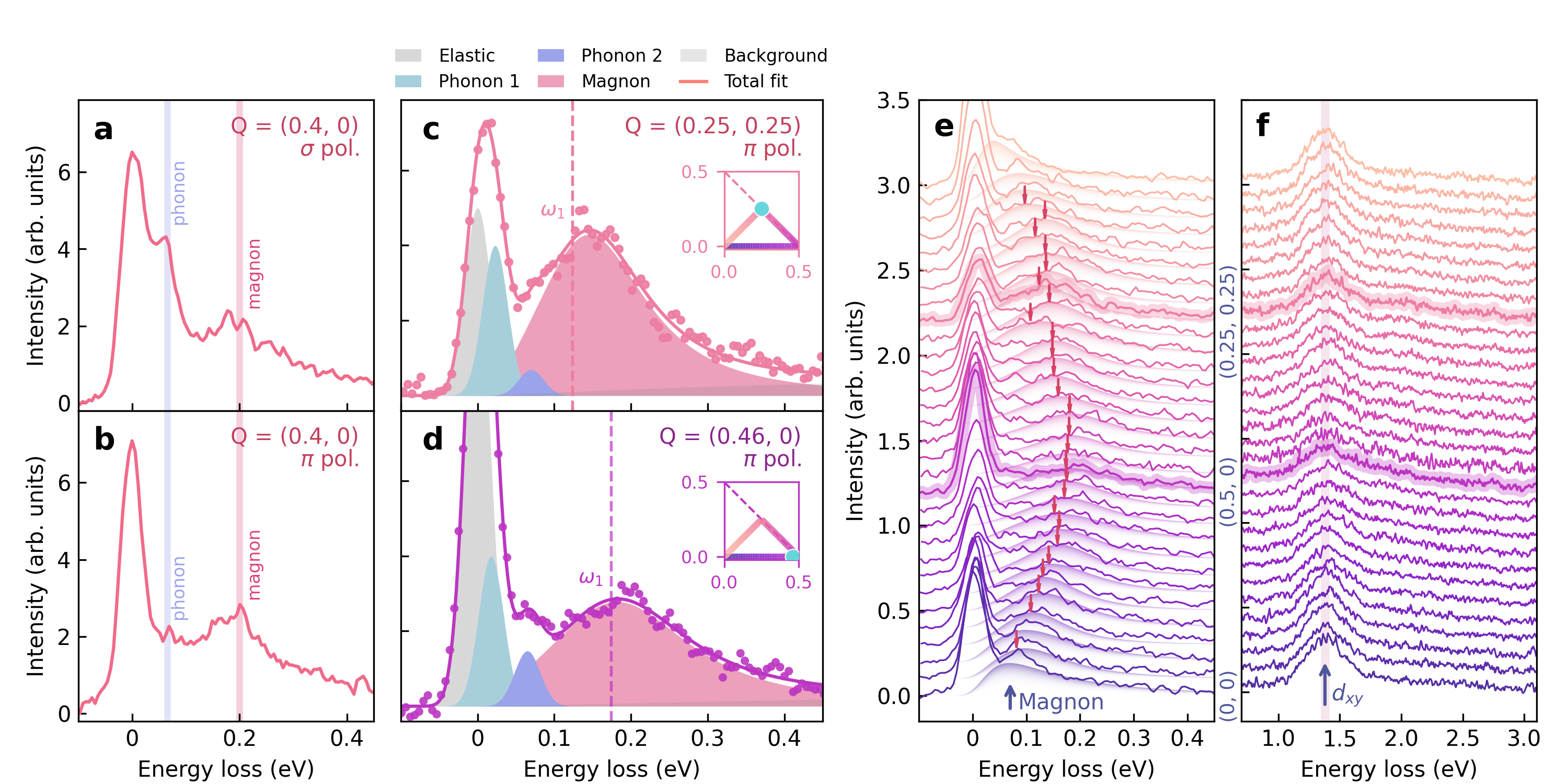}
    \caption{RIXS spectroscopy of infinite layer \PNO. (a,b) Representative RIXS spectra of PrNiO$_2$ with polarization as indicated. Vertical lines mark the phonon and magnon modes. (c,d) RIXS spectra recorded with $\pi$ polarization on PrNiO$_2$ at two high-symmetry zone boundary points as depicted in the insets. The modeling fits are described in the text. Grey component covers elastic scattering and the two blue components are interpreted as phonon excitations. Red shaded areas models the magnetic excitations. Vertical dashed lines indicate the obtained poles. (e,f) RIXS spectra of \PNO\ along three high-symmetry directions. Red arrows in (e) indicate the dispersion of damped magnon energy $\omega_1$ (see Supplementary Information) of the fits. 
    Vertical line in (f) is a guide to the eye.} 
    \label{fig:enter-label}
\end{figure*}

\section{Results}
X-ray absorption spectra (XAS) using 
vertical incident light polarization across the nickel and copper $L$-edges --- shown in Fig.~1 --- demonstrate the quality of our \PNO\ and \SCO\ films.
The shoulder in the XAS of \PNO\ about $1$~eV above the main peak has been attributed to self-doped hole states~\cite{rossi_orbital_2021,yan_persistent_2025}.
At grazing-incidence geometry, the absorption peak is enhanced with $\sigma$ incident X-rays, demonstrating a dominant character of in-plane $d_{x^2-y^2}$ orbitals (see Supplementary Information).
The RIXS energy tuning map --- shown in Fig.~1b --- reveals that magnetic and orbital excitations are resonating at the Ni $L$-edge.
RIXS spectra recorded on \SCO\ and PrNiO$_2$ at the Cu and Ni $L$-edges are shown in Fig.~1c and 1d. For both compounds, the \textit{dd} orbital  excitations are the most intense and a weaker magnon excitation is observed near the elastic line. The \textit{dd} excitations are analyzed and indexed according to the $d$-orbital level splitting in a square-planar crystal-field environment for both \SCO\ and PrNiO$_2$.  
Consistent with an earlier report on CaCuO$_2$~\cite{martinelli_collective_2024}, the $d_{xy}$ orbital excitation displays a significant dispersion in \SCO. For \PNO, this collective behavior is not resolvable within the experimental resolution.

In Fig.~2, we focus on the low-energy part of the RIXS spectra recorded near high-symmetry zone-boundary points with $\sigma$ and $\pi$ polarizations. The spectra with $\pi$ polarization are analysed using a four-component model. Gaussian profiles are used for the elastic scattering (grey shaded area), and two phonon modes (blue shades). A damped harmonic oscillator function~\cite{monney_resonant_2016,lamsal_extracting_2016} is applied to model the magnon modes. From the ``pole" of the damped magnon excitation (see Supplementary Information for details), we extract the magnon dispersion --- see Fig.~2e and 3d-f.

\section{Analysis}
Before providing a refined analysis of the magnon dispersions, we emphasize the key observables and their implications. 
We find that the magnon zone boundary dispersion is comparatively stronger in \SCO. 
Similarly, the $d_{xy}$ orbital excitation exhibits a stronger dispersion in \SCO. Within a single-band Hubbard model, the zone-boundary magnon dispersion is dictated by a series of higher-order exchange terms such as $t^4/U^3$, $t^{\prime 2}/U$, and $t^{\prime\prime 2}/U$ which gain prominence when the correlation $U/t$ decreases. 
In a  strong-coupling scenario, the $d_{xy}$ dispersion originates from the orbital superexchange, which roughly
scales with $t^{\prime 2}/U$~\cite{martinelli_collective_2024}. These observations thus suggest higher-order terms are more pronounced in \SCO, leading to the conclusion that \PNO\ is more correlated compared to \SCO. The question is by how much.

\begin{figure*}
    \centering
    \includegraphics[width=0.9\linewidth]{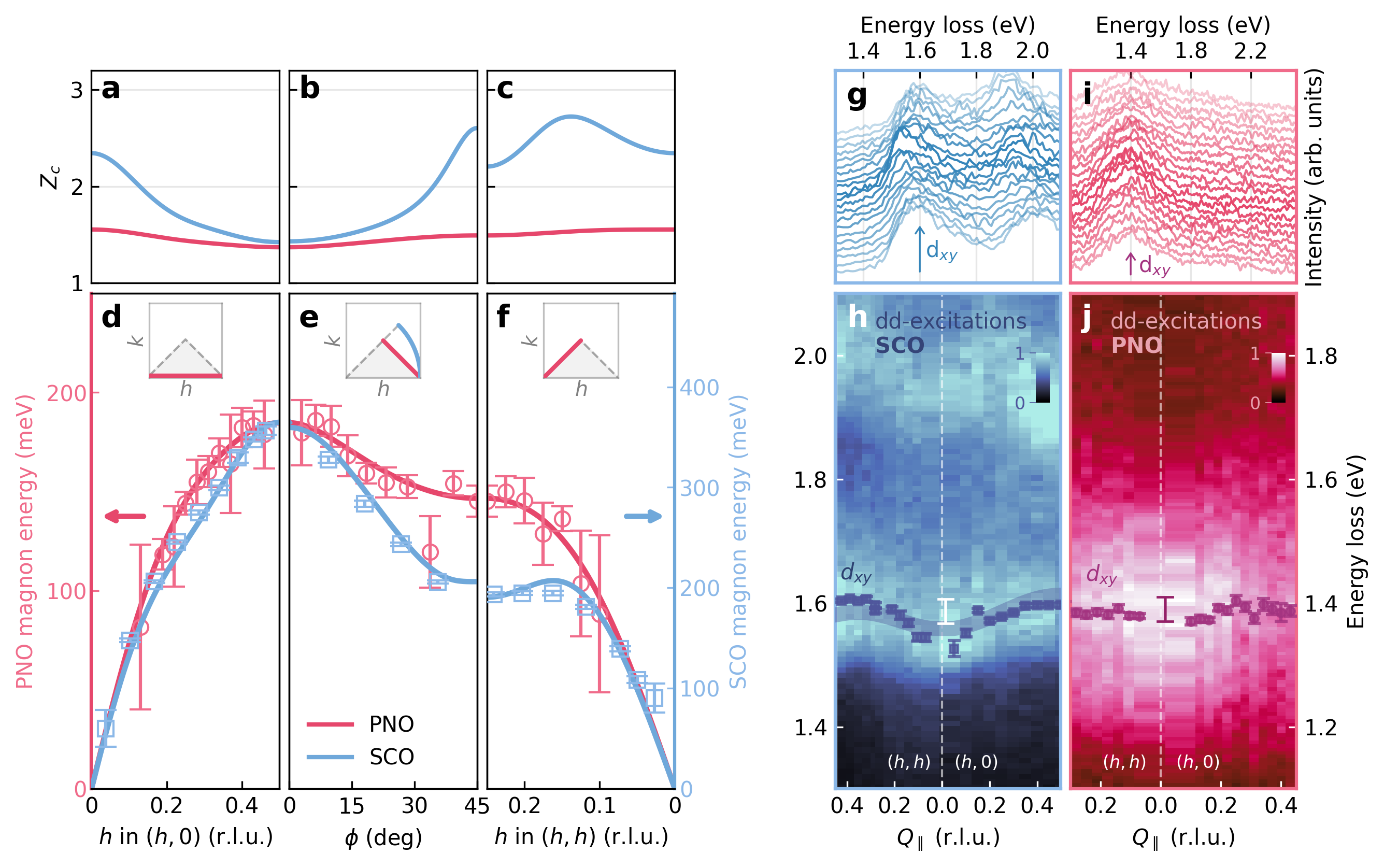}
    \caption{Dispersion of magnons in \PNO\ and \SCO. (a-c) Renormalization factor $Z_c$ for \PNO\ (red) and \SCO\ (blue)  along the momentum trajectories indicated in the corresponding insets of (d-f). (d-f) Dispersions of magnons for \PNO\ (right y-axis, red) and \SCO\ (left y-axis, blue) along the same trajectories. Symbols represent measured data with error bars, and solid lines are fits. Insets in (d-f) show the momentum path in the Brillouin zone for each case. In (b) and (e), the two samples follow different trajectories, as highlighted in the inset. Therefore, we use azimuthal angles $\phi$ to label the momentum transferred. (g,i) RIXS spectra --- recorded along the $(h,0)$ and $(h,h)$ directions for \SCO\ and \PNO\ --- concentrating on the $dd$ excitations. Vertical gray lines are visual guides to the eye. (h,j) Same data as in (g,i) but represented in a false color-map format.  Dots are center of the Gaussian fits of the $d_{xy}$ excitations. Vertical bars indicate the experimental energy resolutions. }\label{fig:enter-label}
    
\end{figure*}

Having extracted the magnon dispersion of \PNO\ and compared it with isostructural \SCO, we now turn to discuss the effective spin exchange interactions. Magnon excitations in  cuprates have previously been analysed using a single-band Hubbard model projected into a Heisenberg Hamiltonian~\cite{coldea_spin_2001,headings_anomalous_2010,delannoy_low-energy_2009,dalla_piazza_unified_2012,ivashko_strain-engineering_2019}. 
These studies include analysis of La$_2$CuO$_4$, that on the energy scales of $U$ and $4t$, is  not strictly a single-band system~\cite{matt_direct_2018,horio_two-dimensional_2018}. We are therefore extending the analysis to \PNO, which, like La$_2$CuO$_4$, hosts both dominant $d_{x^2-y^2}$ and other orbital states at the Fermi level~\cite{li_observation_2025}. 
Generally, exchange interactions are parameterized through the Coulomb interaction $U$ and nearest neighbor hoppings $t$, $t^\prime$ and $t^{\prime\prime}$. 
Given the isostructure of \SCO\ and \PNO, we here assume that ratios of square lattice hopping parameters $t'/t$ and $t^{\prime\prime}/t'$ are constants.
An identical and significant $t'/t$ ratio has been suggested for infinite-layer nickelates and cuprates by first-principles calculations~\cite{botana_similarities_2020}.
Specifically, we use the hopping parameters found for  cuprates --- that is, $t'/t=-0.4$~\cite{matt_direct_2018} and $t^{\prime\prime}/t^\prime=-0.5$~\cite{yoshida_systematic_2006}.
Within this formalism, zone boundary dispersion --- as observed in both \PNO\ and \SCO\ --- implies non-negligible higher-order exchange interaction terms~\cite{coldea_spin_2001,headings_anomalous_2010,delannoy_low-energy_2009,dalla_piazza_unified_2012}. In a recent work on \SCO, it was shown that such high-order terms alone are insufficient in describing the zone boundary dispersion found in \SCO~\cite{wang_magnon_2024}. To account for the observed magnon dispersion, a self-consistent evaluation of the magnon renormalization factor $Z$ is required. This implies that the magnon dispersion is described by the product of momentum dependent pole $\omega_k$ and momentum dependent renormalization factor $Z_k$. Applying this model to \PNO\ and \SCO, it is possible to derive a direct comparison between the nickelates and cuprates. The obtained momentum dependence of $Z_k$ is shown in Fig. 3a-c. The results of our parametrization are given in Table~1. Over all, we find that the Coulomb interaction $U$ is lower (by 25-30\%) in \PNO\ compared to \SCO. However, the correlation parameter $U/t$ is about 20\% stronger in \PNO.

\begin{table}[!h]
\centering
\setlength{\tabcolsep}{7.6pt}
\begin{tabular}{@{}lcccc@{}}
\toprule
\hline\hline
\textbf{Sample} & $t$ [meV] & $U$ [eV] & $U/t$   & Ref. \\
\midrule
\hline
LCO/STO       & 408 & 3.38 & 8.3  & \cite{wang_magnon_2024}\\
\textbf{PNO/STO}       & \textbf{247}   & \textbf{2.0}  & \textbf{8.1}  &   This work \\
Tl2504  & --- & --- & 6.9 & \cite{bialo_magnetic_2025} \\
CCO       & 500 & 3.39 & 6.8  &\cite{wang_magnon_2024} \\
\midrule
\textbf{SCO/GSO}  & \textbf{431.5} & \textbf{2.68} & \textbf{6.2}  &  This work, \cite{wang_magnon_2024}   \\
\hline
\bottomrule
\end{tabular}
\caption{Electronic parameters $t$ and $U$ for \PNO\ and cuprate systems as indicated. These parameters are extracted from a single-band Hubbard model projected into a Heisenberg Hamiltonian from which the observed magnon dispersion is fitted. For this fit, we assume 
$t'/t=-0.4$ (realistic for cuprates~\cite{matt_direct_2018}) and $t''/t' =-0.5$~\cite{yoshida_systematic_2006}.}
\end{table}

\section{Discussion}
Although RIXS experiments have revealed collective spin excitations that persist across a wide doping range of infinite-layer nickelates~\cite{lu_magnetic_2021,gao_magnetic_2024,yan_persistent_2025}, their intrinsic ground states remain elusive. Current experimental characterizations of undoped compounds suggest either a weakly insulating or superconducting state~\cite{lee_linear_2023,parzyck_superconductivity_2025,sahib_superconductivity_2025}, with some evidence for spin-glass behavior~\cite{lin_universal_2022,ortiz_magnetic_2022,saykin_spin-glass_2025,zhou_origin_2025}. However, it is still unclear whether the absence of long-range magnetic order in bulk \PNO\ arises from self-doping or is influenced by sample synthesis conditions~\cite{li_absence_2020,lin_universal_2022,ortiz_magnetic_2022}.
Recent studies have further complicated the picture: a muon spin rotation ($\mu$SR)~\cite{fowlie_intrisic_2022} and an X-ray magnetic circular dichroism (XMCD) experiment~\cite{krieger_signatures_2024} reported evidence for local magnetic moments from Ni$^+$ ions in thin films, whereas inelastic neutron scattering measurements of bulk LaNiO$_2$ single crystals failed to observe spin fluctuations~\cite{hayashida_lattice_2025}. These conflicting experimental results further complicate the intrinsic magnetic ground state in these systems.

In this context, an estimate of the staggered magnetization $\left\langle M_s \right\rangle$ from the spin excitations would provide insights into this question. For Mott insulating antiferromagnets, the staggered magnetization $\left\langle M_s \right\rangle$ is defined by the effective spin number but can be reduced by charge fluctuations. In La$_2$CuO$_4$, spin-wave calculations 
using a Hubbard model that includes
first to third nearest neighbor hopping integrals gives a $\left\langle M_s \right\rangle$ very close to the value obtained from neutron diffraction measurements~\cite{delannoy_low-energy_2009}. Our recent RIXS study of \SCO\ has shown that further decreasing the correlation strength $U/t$ will disrupt the N\'eel order and induce other magnetic ground states~\cite{wang_magnon_2024}. This analysis thus provides an effective experimental approach to infer the magnetic ground state by analyzing the spin excitations.
From the extracted electronic parameters of \PNO, we obtain $\left\langle M_s \right\rangle\simeq0.117$, which is smaller than $\left\langle M_s \right\rangle \simeq 0.235$ of La$_2$CuO$_4$ but larger than that of \SCO~\cite{wang_magnon_2024}.

The strength of electronic correlations significantly influences unconventional superconductivity. Recent theoretical studies on single-band Hubbard model using dynamical vertex approximation (D$\Gamma$A) and density functional theory (DFT) calculations demonstrated that $T_c$ exhibits a dome-like dependence on $U/t$~\cite{kitatani_optmizing_2023,worm_spin_2024}. Excessively strong electronic correlations suppress superconductivity by enhancing antiferromagnetic fluctuations, which open a pseudogap and weaken the electron propagator. Optimal $T_c$ is achieved at intermediate coupling ($U/t\approx6$-$7$).
The larger $U/t$ in \PNO\ compared to \SCO\ places it outside this optimum, which may account for its lower $T_c$. This aligns with our finding that undoped \PNO\ may host a larger staggered magnetization compared to \SCO. Notably, it has been further suggested that reducing $U/t$ via external pressure or compressive strain can shift nickelates toward the optimal correlation range~\cite{ren_possible_2023,lee_synthesis_2025}. This is consistent with the observation of $T_c$ enhancement from $17$~K to over $30$~K in Pr$_{0.82}$Sr$_{0.18}$NiO$_2$ thin films under hydrostatic pressures~\cite{wang_pressure-iduced_2022}. 
A recent study on cuprates also reveals the critical role of moderate correlation strength in optimizing $T_c$~\cite{bialo_magnetic_2025}. Regardless of the correlation and superconducting pairing strength, both cuprates and infinite-layer nickelates exhibit strange metal behaviour around optimal doping~\cite{hsu_transport_2024,lee_linear_2023,iorio-duval_quantum_2025}. Overall, their normal state electronic properties share common characteristics, suggesting similar essential physics despite the variation in correlation strength.

\section{Methods}
Thin films of the PrNiO$_2$ on SrTiO$_3$ (STO) substrates are prepared following procedures described in Ref.~\cite{gao_magnetic_2024}. 
Ni $L$-edge RIXS experiments were performed at the I21 beamline of the Diamond Light Source (DLS)~\cite{zhou_i21_2022}. The energy resolution, determined by the full-width-at-half-maximum of the elastic scattering profile of a carbon tape, was set to 39 meV. The scattering angle was fixed to 154$^{\circ}$, and the sample temperature was set to 16 K.
Cu $L$-edge RIXS experiments on thin films of SrCuO$_2$ grown on GdScO$_3$ (GSO) substrates were carried out at the ADRESS and ID32 beamlines at respectively the Swiss Light Source (SLS) and the European Synchrotron Radiation facility (ESRF)~\cite{brookes_beamline_2018}. Experiments at ID32 were performed with an energy resolution of 39.5 meV and a scattering angle of 145$^{\circ}$, at a sample temperature of 25 K.
Momentum transfer is indexed in reciprocal lattice units (r.l.u.) of the tetragonal unit cell.
Experiments at the ADRESS beamline have previously been described in Ref.~\cite{wang_magnon_2024}. All RIXS spectra are normalized to the area of the $dd$ excitations (400–4000 meV).\\

\noindent\textit{Note added.} Upon completion of this work, we became aware of an independent RIXS study of magnetic and orbital excitations in infinite-layer nickelate and cuprate films~\cite{rosa_spin_2025}.


\section{Acknowledgements}

We thank Kai-Yuan Qi and Dao-Xin Yao for helpful discussions. 
The work at CUHK is supported by the Research Grants Council of Hong Kong (ECS No. 24306223), and the Guangdong Provincial Quantum Science Strategic Initiative (GDZX2401012). X.H., I.B., L.M. and J.C. thank the Swiss National Science Foundation under Projects No. 200021\_188564. 
Part of this research from IOP is supported by the National Key Research and Development Program of China (Grant No.  2021YFA1401800 and 2022YFA1403900).
We acknowledge the DLS and the ESRF for providing beamtime under Proposals MM30189 and HC-5223 at the I21 and ID32  beamlines.

\end{document}